\documentclass[10pt,twocolumn,letterpaper]{article}

\usepackage{usenix}
\usepackage{sectsty}
\sectionfont{\fontsize{12}{16}\selectfont}
\subsectionfont{\fontsize{10.5}{14}\selectfont}
\subsubsectionfont{\fontsize{10}{10}\selectfont}

\usepackage[labelfont=bf]{caption}    % added 9/10/13

\usepackage{courier}            % standard fixed width font
\usepackage[scaled]{helvet} % see www.ctan.org/get/macros/latex/required/psnfss/psnfss2e.pdf
\usepackage[hyphens]{url}                 % format URLs
\usepackage{listings}          % format code
\usepackage{enumitem}      % adjust spacing in enums
\usepackage[colorlinks=true,allcolors=blue,breaklinks,draft=false]{hyperref}   % hyperlinks, including DOIs and URLs in bibliography
\usepackage{titlesec}
\usepackage[hyphenbreaks]{breakurl}
\usepackage{colortbl}
\usepackage{alltt}
\usepackage{hhline}
\usepackage{multirow}
\usepackage{stfloats}
\usepackage{epsfig}
\usepackage{subcaption}
\usepackage{flushend}
\usepackage{xspace}
\usepackage{cite}
\usepackage{fancyhdr}
\usepackage{booktabs}
\usepackage{amssymb}

\newcommand{\xcomment}[1]{{}}
\newcommand{\para}[1]{\smallskip\noindent {\bf #1} }

\usepackage{color}
\definecolor{LightGray}{gray}{0.85}
\definecolor{VeryLightGray}{gray}{0.98}

\newcommand{\comment}[1]{{}}

\newcolumntype{C}[1]{>{\centering\let\newline\\\arraybackslash\hspace{0pt}}m{#1}}
\newcolumntype{L}[1]{>{\raggedright\let\newline\\\arraybackslash\hspace{0pt}}m{#1}}
\newcolumntype{R}[1]{>{\raggedleft\let\newline\\\arraybackslash\hspace{0pt}}m{#1}}

\newcommand{\comments}[1]{{}}

\newcolumntype{C}[1]{>{\centering\let\newline\\\arraybackslash\hspace{0pt}}m{#1}}

\titlespacing*{\section}
{0pt}{1.8ex plus 1ex minus .1ex}{1.0ex plus .1ex}
\titlespacing*{\subsection}
{0pt}{1.5ex plus 1ex minus .1ex}{1.0ex plus .1ex}
\titlespacing*{\subsubsection}
{0pt}{1.2ex plus 0.8ex minus .08ex}{0.8ex plus .08ex}

\begin{document}

\title{{\vspace{-18pt} {\bf Configuration Testing}\\
{\Large Testing Configuration Values as Code and with Code}}}

\author{Tianyin Xu and Owolabi Legunsen\\
University of Illinois at Urbana-Champaign\\
% \{tyxu, legunse2\}@illinois.edu
}
\date{}

\maketitle

\begin{abstract}
\vspace{1.5pt}

\noindent
This paper proposes {\it configuration testing}---evaluating
  configuration values (to be deployed)
  by exercising the code that uses the values and
  assessing the corresponding program behavior.
We advocate that configuration values should be systematically {\it tested}
  like software code
  and that configuration testing should be a key reliability
  engineering practice for preventing misconfigurations from production deployment.
%  one major root cause of failures and outages
%  of today's large-scale production systems.

The essential advantage of configuration testing is to
    put the configuration values (to be deployed)
    in the context of the target software program under test.
In this way, the dynamic effects of configuration values
    and the impact of configuration changes
    can be observed during testing.
Configuration testing overcomes the fundamental limitations of
  {\it de facto} approaches to combatting misconfigurations,
  namely configuration validation and software testing---the former
  is disconnected from
  code logic and semantics, while the latter
  can hardly cover all possible configuration values and their combinations.
Our preliminary results show the effectiveness of configuration testing
    in capturing real-world misconfigurations.

We present the principles of writing new configuration tests
  and the promises of retrofitting existing software tests
  to be configuration tests.
We discuss new adequacy and quality metrics for configuration testing.
We also explore regression testing techniques
  to enable incremental configuration testing during continuous
  integration and deployment in modern software systems.
\end{abstract}

\section{Introduction}
\label{sec:introduction}

In large-scale, rapidly-evolving software systems,
  software configurations are changed frequently~\cite{maurer:15}.
Software engineers constantly change configuration values to
  customize the runtime behavior of production systems.
For example, Facebook reports that ``configuration diffs'' (changes
  to configuration files) are committed
  thousands of times
  a day, more frequently than code changes~\cite{tang:15}.

The velocity of configuration changes makes misconfiguration a significant
  threat to the correctness, reliability, and security of production systems.
Despite the common practice of ``configuration as code'' which enforces
  rigorous quality assurance (including diff review, validation,
  and canary analysis),
  misconfigurations are still
  among the major causes of system failures
  and service incidents of today's cloud and Internet services,
  as reported by numerous failure studies and news reports~\cite{barroso2018,
  xu:15,
  yin:11,
  kendrick:12,
  gunawi:16,
  maurer:15,
  oppenheimer:03,
  nagaraja:04}.
In our experience, misconfigurations that
  cause real-world
  system failures
  are typically not trivial mistakes
    (e.g., typos) but sophisticated ones that
    violates subtle constraints and thus
    are hard to spot via diff
    reviews or captured by rule-based validation.
Moreover, the offending configurations
  are often not absolutely wrong,
  but lead to undesired program behavior;
  sometimes, even valid configuration changes could trigger
  dormant software bugs.

We argue that testing is one essential missing piece
  in today's configuration management practice.
Testing can overcome the fundamental limitations
  of existing configuration validation: being disconnected
  from the program logic and semantics (cf.~\S\ref{sec:state}).
Despite being treated as code,
  configurations are not being tested like code.
While configuration values cannot be directly executed and
  tested on their own,
  the values can be and should be tested
  {\it together with the code}, in order to exercise
  the semantics and observe the dynamic effects of the values.

We use the term\footnote{In the past,
  the term ``configuration testing'' was
  used to refer to {\it configuration-aware software testing}~\cite{Mukelabai:2018}
  and {\it compatibility testing}~\cite{perry:16}, which completely differs
  from its literal meaning (evaluating the correctness of configuration values)
  used in this paper.}, {\it configuration testing},
  to describe the proposed testing effort of
  {\it evaluating configuration values by
      exercising the code that uses the values
      and assessing the corresponding runtime behavior.}
% In configuration testing,
%  the correctness of a configuration value is defined by the behavior
%  of the code using the value.
The essential advantage of configuration testing is to
      put the configuration values (to be deployed)
      in the context of the target software program under test.
  In this way, the dynamic effects of configuration values
      and the impact of configuration changes
      can be observed during testing.

From the testing perspective, a configuration value is not too different from a
  constant value, once the value is {\it configured} (fixed).
Traditional software testing can evaluate constant values in code,
  but cannot effectively deal with configurable values due to the difficulties
  of covering all possible values and their combinations.
Configuration testing does not attempt to cover all possible values,
but focuses on the values to be deployed to production.
A configuration test plugs the configured values (to be deployed) in the test code,
and evaluates the values based on the behavior of the code using the values.
We show a concrete configuration test in \S\ref{sec:testing} (Figure~\ref{fig:example}).

\begin{figure*}
    \centering
    \includegraphics[width=0.99\textwidth]{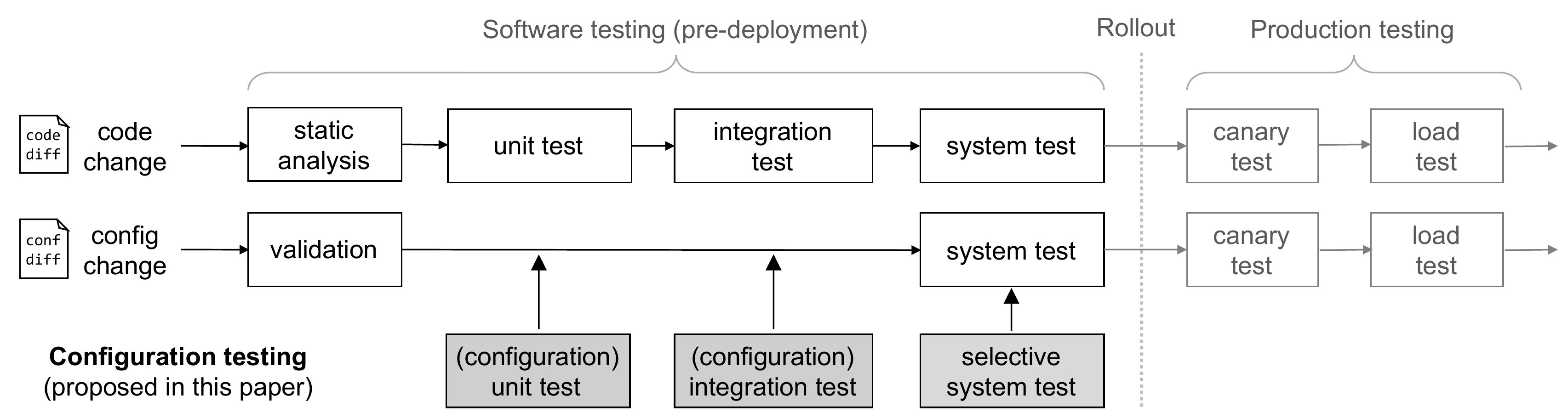}
    \vspace{5pt}
    \caption{The position of {\it configuration testing}
      in the state-of-the-art
      configuration management and deployment process, in comparison to
      code changes~\cite{tang:15,perry:16}.
      In this paper, we mainly focus on testing at the unit and the integration levels
      before production deployment; the same idea and principles can be
      applied to system-level testing as well.
    }
    \label{fig:position}
\end{figure*}

Configuration testing can be done at the unit, the integration, and the system levels.
The testing can evaluate different properties of the system,
  including correctness, performance, and security.
% we focus on unit- and integration-level testing in this paper~\cite{bland:14,Wacker:15}.
Configuration testing can be done {\it incrementally},
  which tests only the changed configuration values
  using regression testing techniques,
Figure~\ref{fig:position} positions configuration testing in the state-of-the-art
    configuration management and deployment process,
    and compares it with traditional software testing.

%\begin{itemize}
%  \item {\it Validation} can hardly cover all the constraints required by
%    the software at runtime,
%    nor can it capture configuration changes that expose software bugs.
%    {\it The limitation roots in
%    the separation between rule-based validation code
%    and the actual software code that uses configuration values.}
%  \item {\it System and canary tests} are expensive and can only serve
%    as the {\it last} levels of defense,.
%    In addition, these tests may not reach every usage of configuration values
%    in the software, especially for configurations
%    used only under special conditions, such as error handling and failure
%    recovery~\cite{xu:16}.
%\end{itemize}

\subsection{Practicality}
\label{sec:practicality}

We believe that configuration testing is practical with few
  barriers to adoption.
With modern reliability engineering practices and
  the DevOps movement~\cite{tang:15,maurer:15,sayagh18,sre_workbook:18} as
  well as the significant impact of configuration changes,
  configuration management has
  already been
  done in a systematic way (the configuration-as-code practice~\cite{tang:15,maurer:15}).
This sets up the natural framework and process for configuration testing,
  as shown in Figure~\ref{fig:position}.

DevOps breaks the longstanding assumption that configurations are managed by
  traditional system administrators (sysadmins) portrayed as those who do
  not read or write code, and do not understand a system's
  internal implementation~\cite{xu:13,moskowitz:15}.
In the era of configuration as code, configurations are managed by engineers
  who implement the software and test their code continuously.

Specifically, we will show in \S\ref{sec:transform} that
  many existing software tests
  naturally include the test logic for configurations,
  which indicates that configuration tests can be implemented and
  maintained like existing software tests.
Note that configuration testing can also be applied in the
   traditional sysadmin-based settings.
   It requires software developers to
   implement configuration tests and release them to sysadmins.

Configuration testing can be run in hermetic environments~\cite{narla:12},
  canary services~\cite{Davidovic:18},
  or actual deployments, similar to existing validation/testing practices~\cite{sre_workbook:18,narla:12,perry:16}.
  % because the configurations
  % could point to environmental objects, e.g., files, IP addresses,
  % and environment variables.
  % This is the same requirement for running validation code
  % rather than a new one.
%With the use of container and virtual machine technologies,
%  high-fidelity testing environment can be effectively
%  created~\cite{sre_workbook:18,xu:18,narla:12,perry:16}.
  % the deployment environment can be efficiently created for testing
  % using the
  % same container/VM images~\cite{sre_workbook:18,xu:18,perry:16}.

\subsection{Testing Framework and Tools}

Configuration testing can be directly supported by existing
  software testing techniques.
Configuration testing can be run on top of existing testing
  infrastructure.
In principle,
  both configuration testing and traditional software testing exercise
  the code under test and assert the expected behavior (e.g., program outputs).
Configuration tests can be implemented using existing test frameworks
  such as JUnit for unit-level configuration tests,
  as demonstrated in Figure~\ref{fig:example}.

% The difference lies in the testing purposes.
% Traditional software tests use
%  a few representative configuration values.
% The representative values hardcoded in tests are typically the
%  commonly-used configuration values,
%  as the purpose is to examine the correctness of code implementation.
% Configuration testing, on the other hand, tests the actual {\it configured} values
%  to be deployed
%  by evaluating whether the code under test
%  can use these values to deliver the expected program behavior.

We discuss tooling support for configuration testing,
  including test generation, test adequacy measurement,
  and test selection for incremental testing.
Specifically, we observe that
  many existing software tests can be reused and retrofitted into
  configuration tests using the parameterization-and-concretization transformation:
  (1) parameterizing hardcoded configuration values in the test code,
  and (2) concretizing the parameterized value with the actual configured values
  to be deployed to production.
We discuss the feasibility and promises of automatically
  generating configuration tests,
  and the challenges of evaluating the quality of auto-generated
  configuration tests.
We also discuss the techniques and metrics for measuring test adequacy of
  configuration test suites.
Last but no the least, we discuss test selection to enable
  incremental configuration testing
  in the context of continuous integration and deployment~\cite{Savor:16,Rossi:16}.

\section{Limitations of the State of the Art}
\label{sec:state}

In order to establish the context necessary to understand the advantages of
  configuration testing,
  we discuss the fundamental
  limitations of the state-of-the-art research and practices
  for combatting misconfigurations, including
  configuration validation and system tests.
% We focus our discussion on configuration validation
%  as it is the primary {\it pre-deployment} defense.

% This section reflects on the fundamental limitations of
%  the state-of-the-art research
%  and practices for defending against misconfigurations.
% The reflection drives the motivation of configuration testing.

\subsection{Validation cannot replace testing}
\label{sec:validation}

Configuration validation checks configuration values using validation code
  written by software engineers.
The validation code checks configuration values based on predefined
correctness rules regarding the expected data type, data range, data format, etc.

We argue that rule-based validation should not be the primary quality assurance
  for configuration changes,
  in response to the recent trend of investing in extensive validation~\cite{Baset:2017,tang:15,huang:15,
  santo:16,santo:17,Tuncer2018,liao:18,zhang:14,potharaju:15,Tuncer2018}.\footnote{{\it Validation}
      is referred to as checking configuration values based on
      external specifications.
      It is different from {\it testing} that
      evaluates how configuration values are
      internally used by the system~\cite{Adrion:1982}.}
While rule-based validation can provide basic sanity checks,
it suffers from a number of fundamental limitations:
\begin{itemize}
  \item {\it Configuration value validation is disconnected from the code logic
    and semantics, and thus
    cannot capture undesired program behavior induced by configuration values.}
    Essentially, the predefined correctness rules are based on
      external specifications of the values, and thus are
      completely agnostic to the program behavior under the values.
    As a result, existing configuration validation cannot combat
      {\it legal misconfigurations} which have valid values
        (satisfying the specifications)
    but do not deliver the desired program behavior.
    As reported in recent studies~\cite{yin:11,tang:15},
      legal misconfigurations contribute to a large portion (46.3\%--61.9\%) of
      real-world misconfigurations that caused production impact.

\item {\it Validating configuration values alone cannot combat {\it valid}
    configuration changes triggering dormant code bugs.}
    As reported in Facebook's study~\cite{tang:15},
      among the configuration-induced incidents,
      22\% of them were caused by valid configuration changes
      that trigger software bugs.
      In addition, the validation rules derived from external specifications
        often do not
        match the constraints required by the actual
        implementation due to software bugs~\cite{Adrion:1982}.
    One common pattern is
          misinterpretation of raw configuration values when
          parsing them from files due to
          undefined specifications
          %~\cite{hdfs-5262,hdfs-5690}
          and bugs in code.

  %~\cite{hadoop-7721,hdfs-2683,hdfs-3439}.}
  %  Configuration changes that expose software bugs have valid values---the
  %    root cause is in the code affected by the changes.

  \item {\it It is prohibitively difficult and expensive to manually codify
    the complete rule set for every single configuration parameter.}
    Our prior work~\cite{xu:16} shows that one configuration value could
    have multiple different constraints, and constraints could be subtle and
    hard to codify into static rules.
    Prior work proposes to {\it automatically} infer
    constraints from field configuration data~\cite{santo:16,santo:17,zhang:14,yuan:11,Tuncer2018},
    documents~\cite{potharaju:15},
    and source code~\cite{xu:13,liao:18,rabkin:11}.
    However, all those methods can only infer a few
    specific types of constraints.
\end{itemize}
Figure~\ref{fig:val_vs_test} illustrates the difference in principles
  of configuration validation versus configuration testing.
In our viewpoint, validation and learning-based methods (\S\ref{sec:data_driven})
  should supplement configuration testing by
  checking empirically good practices and hidden patterns, or being used
  when configuration tests are unavailable.

% \vspace{3pt}
% \para{\it  Validation is not cheap.}Validation requires
%  implementing and maintaining validation code, typically
%  using separate languages and frameworks~\cite{Baset:2017,tang:15,huang:15,zhang:14}.
% Maintaining validation code requires excessive and continuous
% effort, because: 1) modern
% software systems expose enormous numbers of configuration parameters~\cite{xu:15:2};
% 2) a parameter may have multiple constraints;
% and 3) constraints change with software evolution~\cite{zhang2:14}.

% \vspace{3pt}
%\para{\it  Empirical evidence shows that configuration validation is
%  often deficient even with mature
%   engineering practices}~\cite{xu:16,tang:15}.
% Our prior study reveals that 4.7\%--38.6\% of the configurations
% used in error handling and fault tolerance
% do not have checking code~\cite{xu:16}.
  %, leaving the systems vulnerable to latent configuration errors~\cite{xu:16}.
% As reported in Facebook's study~\cite{tang:15},
%  despite extensive validation and canarying,
%  erroneous configuration changes still found their way into production and
%  resulted in service-level incidents, including both obvious and sophisticated
%  misconfigurations.
%  % typos, out-of-bound values, and referencing incorrect clusters;

% Therefore, instead of trying to codify every single constraint
%    into rules,
%    one should directly test how configuration values are used by the code,
%    i.e., the configuration testing proposed in this paper.
% This not only saves the effort of codifying and maintaining validation rules,
%    but also covers precise and complete constraints.

\subsection{Learning-based methods are not magic}
\label{sec:data_driven}

A frequently explored direction is to use machine learning techniques
  to automatically classify correct versus erroneous
  configurations based on learning ``big'' configuration data.
Such learning based methods suffer from the same set of limitations as rule-based
  validation discussed in \S\ref{sec:validation}.
In fact, our experience shows that
    learning algorithms hardly work without being scoped with
    predefined rule templates~\cite{zhang:14}.
Nevertheless, few learning-based methods have guarantee on false positives
  or negatives.
As observed at IBM, without explicit guarantee,
  ``learning-based methods have rarely
  found use in production systems on a continuous basis~\cite{Baset:2017}.''
Specifically:

  \begin{figure}
      %\centering
      \includegraphics[width=0.5\textwidth]{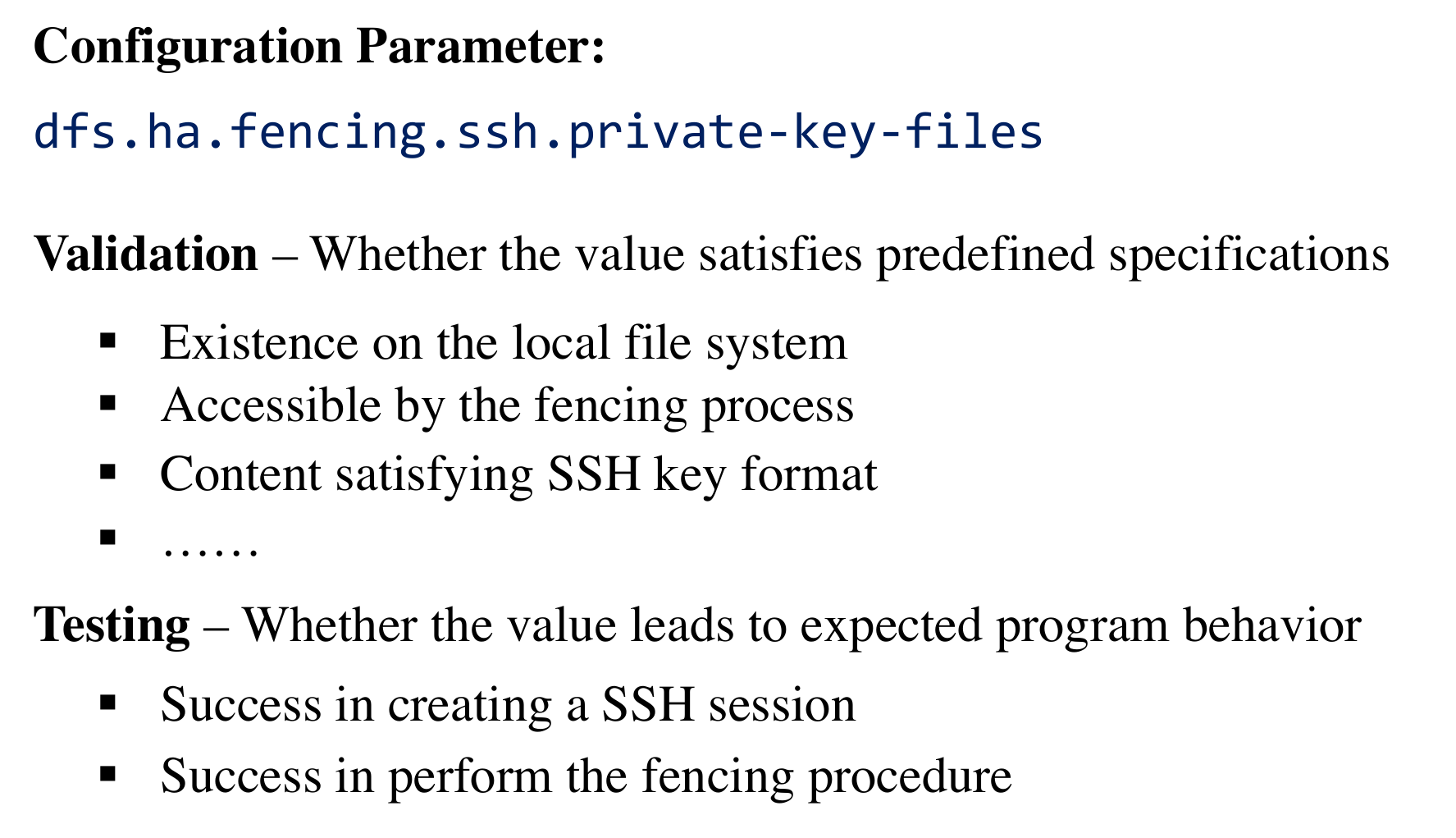}
      \caption{The difference in principles of configuration validation versus
        configuration testing.
      }
      \label{fig:val_vs_test}
  \end{figure}

\begin{itemize}
\item{\it  Misconfigurations may not be outliers and vice versa.}
It is challenging to determine the correctness of
configurations based on their values.
  % without
  % taking software behavior into consideration.
Prior work proposes to use outlier detection algorithms to
detect misconfigurations~\cite{palatin:06}; however, outliers could come
from special customization instead of misconfigurations.
{\it Misconfigurations may not be outliers either.}
Default values are often the mostly-used values, but
staying with defaults incorrectly is a common pattern of misconfigurations~\cite{xu:15:2}.
%Unfortunately, runtime information (e.g., logs and counters)
%is often unavailable before deployment.
% \footnote{The proposed
% configuration testing could help collect runtime logs and traces.}

\item{\it  Datasets are not always available.}
Learning relies on large configuration datasets
collected from {\it independent} sources.
Such datasets are not always available
in typical cloud settings
where all the configurations are managed by the same
  operation team.
Learning-based methods are more suitable for
end-user software with large user bases, e.g., Windows-based applications~\cite{wang:03,
  wang:04,kiciman:04,yuan:06,kushman:10}.
\end{itemize}

\subsection{System tests are not targeted}
\label{sec:not_target}

System tests are
  large-scale tests designed
  for evaluating end-to-end system behavior, only done via
  canary analysis services~\cite{Davidovic:18}.
System tests are expensive and are hard to cover every configuration
  usage.
For a configuration change,
  it is hard to identify whether a test run
    exercised the changed
    configuration values
    and how to measure the impact of the changes.
Furthermore, our prior study~\cite{xu:16} shows that
  configurations can be {\it only}
  used under special conditions; therefore,
  testing steady states
  may not expose latent configuration errors.
% Test selection also becomes difficult---even for
%  a small change that updates one configuration value,
%  the common practice is to run all the system tests.

  % We show in~\S\ref{sec:results} that configuration testing
  % naturally covers
  % most of the rules targeted by the prior work.
  % We discuss what rules will be useful with the configuration testing in place
  % in \S\ref{sec:useful_rules}.

%  The former uses synthetic loads,
%    while the latter deploys new configurations on a
%    subset of production systems that serve live traffic.

%  failing to pass
%  system tests could block deployment,
%  while failures manifested during canary affect the actual production systems.

%  System tests and canary are designed to expose
%  sophisticated misbehavior; one should not abusively rely on them
%  to detect the misconfigurations which
%  can be captured through software testing.

% 1. external testing by workload, hard to guarantee the coverage

\section{Configuration Testing}
\label{sec:testing}

% We advocate that
% configurations should be thoroughly tested at the same level as software code,
% and should be tested together with software code beyond
%%  the state of the art (cf. \S\ref{sec:state}).

The high-level idea of configuration testing
  is to test configuration values by executing the code that
  uses these values and asserting the expected behavior of the code
  (e.g., program outputs).
Unlike traditional software tests that use
  hardcoded configuration values (for the purpose of finding bugs in the code),
  configuration testing exercises software programs with the actual
  configured values to be deployed in production.
Figure~\ref{fig:example} shows a unit-level configuration test, and compares it
  with a unit test shipped with the software project.

From the perspective of configuration testing, a configuration value is not
  essentially
  different from a constant value, once the value is {\it configured} (fixed).
Traditional software testing is able to evaluate constant values in code,
  but cannot effectively deal with configurable values mainly due to
  the challenges of covering all possible
  values and their combinations that may occur in the field.
Configuration testing does not attempt to explore the entire configuration
  space.
Instead, it {\it concretizes} the configurable values with the actual
  configured
  values in the test code,
  and evaluate whether the software using the value
  behaves as expected.
% As discussed in \S\ref{sec:transform}, many existing software tests can be
%  transformed into configuration tests.

Configuration testing can be done at the unit, the integration, and the
  system level to evaluate different scopes
    of the software system under test.
Configuration testing can also be done {\it incrementally} to
  test only the configuration values changed in a given diff.

\subsection{Reusing existing software tests}
\label{sec:transform}

We find that many existing software tests,
  including unit, integration, and system tests,
  can be reused for configuration testing.
  This section focuses on unit and integration tests,
  but the ideas also apply to system tests.

  Conceptually, reusing existing test code for configuration testing involves
    two steps: (1) {\it parameterizing} hardcoded configuration
    values in the test code, and (2) {\it concretizing} the parameterized
    value with the actual configured value to be deployed into production.

  \begin{figure}
    %\centering
    \includegraphics[width=0.505\textwidth]{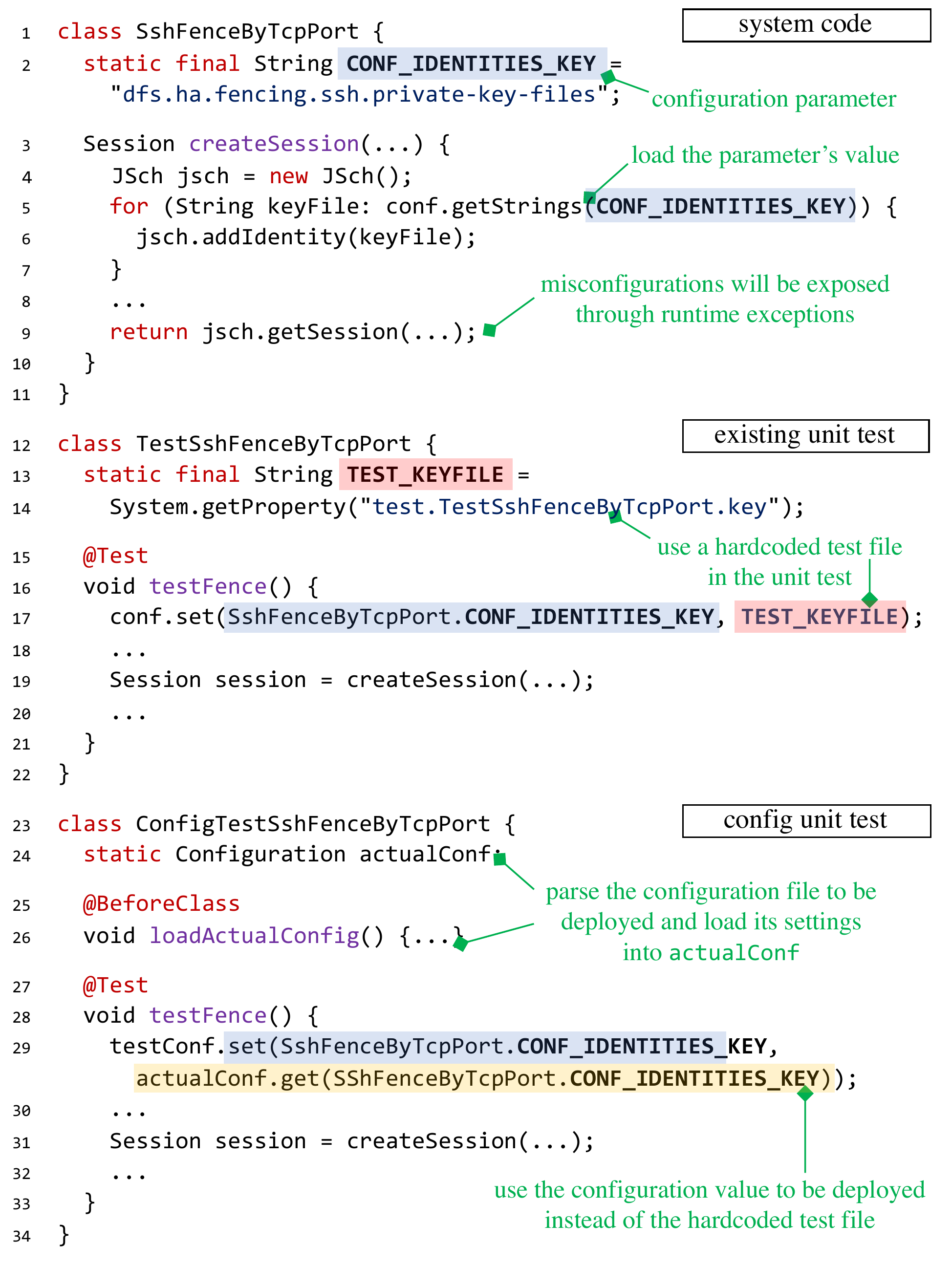}
    \vspace{-18pt}
    \caption{\footnotesize {\small \bf An example of a unit-level configuration test
      versus a traditional
      unit test (using the JUnit framework).}
      Misconfigurations of the parameter would break HDFS NameNode auto-failover
      during a procedure called \texttt{Fence}~\cite{hdfs-7727}.
      The existing unit test
      uses a test file to exercise \texttt{Fence}.
      The configuration test replaces the test configuration (stored
        in \texttt{testConf})
      with the configured value to be deployed (stored in \texttt{actualConf}).
      Misconfigurations will fail
      the test via exceptions thrown at Line 9.
      Validating the configuration is tedious due to
        multiple constraints (existence, accessibility, and file content)---currently HDFS
        does not validate any of these.
      All the constraints can be checked by testing if the configured value can be
      used to create a \texttt{JSch} session and do \texttt{Fence} successfully.
    }
    \label{fig:example}
  \end{figure}

In our experience, the two steps can be {\it systematically}
  done based on well-defined configuration APIs
    in modern software
  systems.\footnote{The observation
  % that modern software systems use well-defined, unified
  %  configuration APIs
  has been validated by many studies on
      real-world systems~\cite{xu:13,tang:15,xu:16,sayagh18}.
% including both commercial and open-source systems~\cite{xu:13,tang:15,xu:16,sayagh18}.
    Typically, mature software projects have customized configuration APIs
    wrapping around common libraries such as \texttt{java.util.Properties}
    for Java in Hadoop, \texttt{configparser} for Python in OpenStack,
    and Thrift structures in Configerator~\cite{tang:15}.
    }
In Figure~\ref{fig:example}, HDFS uses a set of
    \texttt{get} and \texttt{set} APIs for retrieving and rewriting
    configuration values
    stored in a \texttt{Configuration} object. By replacing
    \texttt{TEST\_KEYFILE} with the actual
    configured values (Line 29),
    we transform
    the original software test into a configuration test.
Further, one can {\it automatically} transform existing
  software tests into
  configuration tests by
  rewriting configuration objects or
  intercepting configuration API calls.
Note that such automated transformation may not lead to valid configuration
  tests if the test logic is specific to the original hardcoded values,
  which in our experience is not uncommon.

  \begin{table}[t]
    \footnotesize
    \centering
    \begin{tabular}{L{15mm}C{15mm}C{15mm}C{15mm}}
      \toprule
      \multirow{2}{*}{\bf Software} & \multirow{2}{*}{\bf All Tests} & \multicolumn{2}{c}{\bf Tests Using Configurations}\\
      \cline{3-4}
                 &                 & {\bf \texttt{set}} & {\bf \texttt{get}-only} \\
      \midrule
    %    Hadoop       &  2557 & 337 & 304 & 337 \\
      HDFS         &  3069 & 763 & 559 \\
      YARN         &  2620 & 672 & 229 \\
      MapReduce    &  1302 & 398 & 261 \\
      HBase        &  4977 & 809 & 413 \\
      \bottomrule
      \end{tabular}
      \vspace{-2pt}
      \caption{{\bf Existing software tests that use configuration values.}
        {\small ``\texttt{set}'' refers to tests that explicitly
          customize specific configuration values
          in test code, while ``\texttt{get}-only'' refers to tests that use
          any values in the \texttt{Configuration} objects passed to the test
          (which tends to be generic
          to any values).}
      }
      \label{tab:unit_test_count}
  \end{table}

  \begin{table}[t]
    \footnotesize
    \centering
    \begin{tabular}{L{17mm}C{17mm}C{25mm}}
      \toprule
      \multirow{2}{*}{\bf Software} & {\bf \# Total} & {\bf \# Parameters} \\
                                    & {\bf Parameters}   & {\bf Exercised in Tests}  \\
      \midrule
  %    Hadoop      &  100 & 50 (96.4\%)  \\
      HDFS         &  387 & 373 (96.4\%) \\
      YARN         &  340 & 319 (98.5\%) \\
      MapReduce    &  330 & 335 (96.7\%) \\
      HBase        &  761 & 698 (91.9\%)  \\
      \bottomrule
    \end{tabular}
      \vspace{-2pt}
    \caption{{\bf The number (percentage) of configuration parameters
      exercised by existing software tests.}}
      % The numbers are collected by instrumenting configuration \texttt{getter}
      % APIs and running all the tests.}
      \vspace{-5pt}
      \label{tab:unit_test_coverage}
  \end{table}

%\footnote{We have built
%  a prototype for Java-based systems listed in
%  Table~\ref{tab:unit_test_count} and~\ref{tab:unit_test_coverage}
%  as a proof-of-concept and
%  validated its feasibility.}

\vspace{3pt}
\para{\it Opportunities of reusing existing software
  tests for configuration testing.}To investigate the feasibility of reusing existing tests,
  we analyze the
  test code of four widely-used open-source software projects.
  All of these projects implement unit
  and integration tests using JUnit
  and use configuration APIs similar to those presented
  in Figure~\ref{fig:example}.

Table~\ref{tab:unit_test_count} shows that a significant number of existing tests
  use configuration values in test code.
We build static analysis on top of the Soot compiler framework to analyze
  the test code.
Our static analysis shows that
  these tests create \texttt{Configuration}
  objects and pass them to the code under test.
Therefore, these tests can be potentially reused for configuration testing.
In particular, as shown in Table~\ref{tab:unit_test_count},
  a significant number of these tests do not customize any configuration values
  in the test code---these tests do not \texttt{set} any specific values,
  but only \texttt{get} the default values
  stored in the \texttt{Configuration} object.
We observe that these tests tend to be generic.
The tests
  are supposed to work with any configuration values stored
  in the \texttt{Configuration} object---changing the default
  values should not need to change the test logic.

Table~\ref{tab:unit_test_coverage} shows that
  90+\% of the configuration parameters
  are used by running existing tests.
We instrument configuration \texttt{get} APIs to log
  the configuration parameters retrieved at runtime during the execution
  of the test suite and count the unique parameters in the log---all
  the studied systems retrieve configuration values on demand (when
  they need to use the values).
Note that the numbers do not reflect the coverage metric based on the
  slice of a configuration value defined in \S\ref{sec:coverage}.

%  we instrument the APIs and count the parameters after running all the tests.
% \TODO{What parameters are not exercised?}

\vspace{3pt}
\para{\it Preliminary results.}We evaluate the effectiveness of configuration testing using
  45 latent misconfigurations in the dataset of our prior work~\cite{xu:16}
  for the systems listed in
  Tables~\ref{tab:unit_test_count} and~\ref{tab:unit_test_coverage}.
{\it We find that all the evaluated latent misconfigurations
  can be captured by unit-level configuration testing.}
{\it Most importantly, we find that the configuration tests that are able to
  catch these misconfigurations can be directly created by
  reusing existing tests shipped with the systems.}

Specifically, 43 out of 45 can be detected by running existing test code with
  automated parameterization-and-plugin transformation
  without any modifications;
  the remaining two require
  additional changes of the original test code (for setting up external
  dependencies).

\subsection{Creating new configuration tests}
\label{sec:creation}

We envision software engineers implementing
  configuration tests in the same way that they create unit or integration tests.
Configuration tests requires test framework support for
  parameterizing configuration values in test code
  and concretizing the parameterized values upon configuration changes.
Such support can be built by extending existing test frameworks
  (e.g., on top of
  parameterized test support in JUnit).

Similar to software tests,
  configuration tests need to be maintained continuously
  to accommodate the software evolution.
  For example, new tests need to be added when new configuration
  parameters are introduced, while existing tests need to be revised when
  the usage of configuration values changes in code.
To assist engineers to create new configuration tests,
  tooling can be built
  to identify and visualize code snippets that use configuration values based on
  existing techniques for tracking configuration values in source
  code~\cite{xu:16,attariyan:12,attariyan:10,zhang:13,rabkin2:11,xu:13}.
% The techniques can be used to
%  measure configuration testing coverage in order to enforce test adequacy
%  and quality (discussed in \S\ref{sec:coverage}).s

Automatic configuration test generation is possible. In fact, it is likely
  a simpler problem compared with traditional
  test generation with the goal of exploring all possible program paths~\cite{Cadar:2013}.
Configuration tests only need to cover program paths related to
  the target configuration values,
  which in our experience only touches a small part of the program
  and does not suffer from path explosion.
An effective approach is to %determine the concrete API arguments to
  enforce configuration-related program paths
  based on satisfiability.

  \subsection{Quality}
  \label{sec:quality}

  We use the term ``quality'' to refer to the correctness and effectiveness
    of configuration test cases,
    measured by the false negatives and false positives.
  The quality of configuration tests, especially those
    automatically transformed from existing tests (\S\ref{sec:transform})
   should be carefully evaluated to make configuration testing effective in practice.

  % Soundness and completeness can be empirically measured by evaluating
  The quality of configuration test suites can be empirically evaluated
    using known good and bad configuration values to measure false positives
    and negatives respectively.
  A useful configuration test should pass the good configuration value
  and fail the bad values.

  On the other hand, collecting a comprehensive set of good and bad configuration
    values turns out to be non-trivial---knowing all the good and bad values
    are equivalent to knowing all the constraints of the configuration.
  Fuzzing and constraint-aware mutation based methods~\cite{xu:13,keller:08,li:18,zhang:15}
    can potentially be applied to
    generate correct configurations and misconfigurations.
  The seed configurations can be collected from historically used values~\cite{maurer:15}
    and community-based data sources~\cite{xu:18}.

  % Completeness can be evaluated using
  %  misconfiguration datasets, either from
  %  or generated based on mutation-based techniques~\cite{xu:13,keller:08}.
  %  The idea of mutation testing~\cite{Jia:11} can be borrowed from traditional
  %    software testing
  %    to evaluate the quality of configuration tests.
  %  Instead of mutating programs, one can mutate the
  %    configuration values to generate both correct values and
  %    faulty values to evaluate soundness and completeness of
  %    configuration tests, respectively.
  %  This can be built on top of constraint-aware misconfiguration injection
  %    techniques~\cite{xu:13,keller:08,li:18}.

    % \TODO{A configuration test should satisfy the following quality criteria:
    % (1) no false positive: if the target value is erroneous,
    % the test always fails; and (2) no false alarm: if the target configuration
    % value is correct, the test always passes.
    % The objective of this task is to develop techniques and tools
    % to test the quality of configuration tests.
    % This enables software vendors to identify ineffective configuration tests and improve test quality.
    % ~\cite{Luo:2014,Whittaker:12,Jia:11}

\subsection{Adequacy}
\label{sec:coverage}

As a type of software testing, configuration testing needs adequacy criteria
  for selecting and evaluating configuration test cases.
% Testing requires coverage metrics
% to measure adequacy for test selection,
% prioritization, and minimization~\cite{Yoo:12}.
We find that code coverage metrics (statement, branch,
  and path~\cite{Zhu:1997}) are not suitable as adequacy criteria
  for configuration testing---high coverage of the entire code base
  is an overkill of configuration testing.

We propose {\it configuration coverage} as an adequacy criterion of
  configuration testing.
At a high level, configuration coverage
  describes whether or not the {\it program slice} of
  the target configuration value
  is covered by the configuration tests.
%  based on how the target configuration is used in the program.

\vspace{3pt}
\para{\it Configuration parameters.}For a configuration test suite,
  a configuration parameter is covered
    if the tests exercise all the execution paths in
    the program slice of the parameter.
The program slice of a configuration parameter can be generated using static
  or dynamic taint analysis that takes the parameter's value as initial taints,
  and propagates taints through data- and control-flow dependencies,
  which is a common practice used in prior work~\cite{xu:16,attariyan:12,attariyan:10,
    zhang:13,rabkin2:11}.
Thin slicing~\cite{sridharan:07} is commonly
  used in practice to avoid over-tainting due to unbounded control-flow
  dependencies, while a broader slice definition~\cite{Weiser:81}
  can be used in configuration testing to expose
  bugs trigged by configuration changes.

\vspace{3pt}
\para{\it Configuration changes.}Given a configuration change, the
  tests should exercise not only the
  changed parameters, but also other parameters that {\it depend
  on} the changed ones.
We define that a parameter $P$ depends on another parameter $Q$,
  if the program slice of $P$ is affected by $Q$'s value.
Common patterns of dependencies include both control- and data-flow dependencies.
For example, $P$ is only used when $Q$ has certain value ($Q$ enables a
  feature and $P$ controls the behavior of the feature), or
  $P$'s value is derived from $Q$'s value.
In both cases, when $Q$'s value is changed, $P$ should also be tested.

Note that configuration coverage only measures the program statements that use
  configuration values.
It does not directly measure all the program behavior influenced by
  a given configuration values.
Metrics that can effectively capture the program behavior affected by
  configuration values are desired.

\subsection{Incremental configuration testing}
\label{sec:incremental}

With continuous integration and deployment,
    configurations evolve in frequent updates that only
    change a small number of configuration values.
    For example, Facebook reported that
    49.5\% of configuration updates are two-line revisions,
    while the size of a configuration file can be kilobytes to megabytes~\cite{tang:15}.

The proposed procedure of incremental configuration testing is in the same vein as
  regression testing in continuous integration and deployment.
Given a configuration change, one should selectively run only
    the tests affected the changed configuration values
    and the values that depend on the changed value
    instead of the entire test suite to reduce cost
    and improve efficiency.
The key to testing incremental configuration changes is to associate
      each test with
      the configuration parameter whose {\it impact} can be evaluated by the test.
This can be done by test selection based on the coverage criteria
  in \S\ref{sec:coverage}.

  % For example,
  % running regular smoke tests~\cite{smoke,smoke4}
  % cannot expose the misconfigurations in Figure~\ref{fig:example} which
  % are only manifested when the primary NameNode fails.

\section{Open Problems}

Despite its promises,
  configuration testing
  faces a number of open problems:

\vspace{3pt}
\para{\it Test reuse.}Automated or semi-automatic methods for
  transforming existing software tests into configuration tests
  can significantly reduce the barrier of adoption and bootstrap configuration
  test suites,
  given that mature software projects all have comprehensive software test suites.
  As discussed in \S\ref{sec:transform}, the major challenge is not about
  parameterizing the hardcoded values in existing tests, but to understand
  and analyze the
  test code logic regarding the configuration values. An effective reusing
  method should be able to differentiate hardcoded values that are specific to the
  test cases versus those that are generic,
  or at least identify (and exclude) test cases specific to the
  hardcoded values.

\vspace{3pt}
\para{\it Test generation.}We believe that automated test generation can be
  done at the level of unit and integration tests, in a similar manner as
  test generation for software code.
The feasibility has already been demonstrated by our prior work, PCheck~\cite{xu:16}---the
  checking code generated by
  PCheck is essentially a test.
On the other hand, the test generated by PCheck is basic and does not incorporate
  much of the semantics derived from the code logic due to its limitation
  of dealing with dependencies and side effects, both of which
  can be addressed by configuration testing. Section~\ref{sec:discussion} gives
  a in-depth, retrospective discussion on this matter.

\vspace{3pt}
\para{\it Dependency analysis.}Dependency analysis is essential
  to effective configuration testing,
  especially
  to test selection for incremental configuration changes
  as discussed in \S\ref{sec:coverage}.
While prior work has investigated methods
  to discover dependencies between
  configuration parameters and their values~\cite{zhang:14,xu:13,santo:17},
  none delivers sound and complete results.
It is perhaps reasonable for developers to encode dependencies when
  introducing new configuration parameters,
  while a thorough understanding of
  various types of configuration dependencies is desired.

\vspace{3pt}
\para{\it Testing performance, security, and resource utilization.}
Most of the discussion in this paper implicitly focuses on correctness from
  the software program's standpoint.
On the other hand, the impact of a configuration change often
  goes beyond correctness
  properties, as configurations could affect performance, security,
  resource utilization as revealed in prior
    studies~\cite{attariyan:12,wang:18,rabkin:13,hoffmann:11,xu:17}.
Configuration testing is not limited to correctness,
  and should be applied to other aspects of software systems as well.
One challenge lies in the impact analysis of configuration changes---unlike
  correctness,
  performance, security, and resource utilization is often not
  straightforward or deterministic to measure.

\vspace{3pt}
\para{\it Testing code changes with deployed configuration.}A natural extension
  to the idea of configuration testing is to run the configuration tests
  for code changes with the deployed configuration values.
Such testing can catch bugs that are not exposed in traditional software
  testing due to the inconsistency
  between the configuration deployed in production
  and the configuration hardcoded in the software tests.
Therefore, the configuration tests can be used for testing both configuration
  and code changes: the former plugs the configuration to be deployed, while
  the latter plugs the configuration already deployed.
Note that the testing pipeline could still be separate
 due to the independence of code and configuration rollout.

\section{Discussion}
\label{sec:discussion}

% As a key component of reliability engineering, configuration management has attracted
%  enormous efforts from various research communities,
%  including systems, software engineering, programming languages, and data
%  mining (as shown by the references).

Given the impact of misconfigurations in real-world applications, especially
  cloud and Internet services~\cite{oppenheimer:03,
  nagaraja:04,yin:11,maurer:15,kendrick:12,rabkin:13,gunawi:16,barroso2018},
  recent effort on tackling misconfigurations has shifted from
reactive methods (troubleshooting)~\cite{wang:04,zhang:13,attariyan:10,
  attariyan:12,su:09,mickens:07,rabkin2:11,whitaker:04} to
proactive methods (validation and error detection)~\cite{xu:13,xu:16,huang:15,
  santo:16,santo:17,tang:15,Baset:2017,potharaju:15,zhang:14}.
Configuration testing is along this line, aiming at proactively
  capturing undesired system behavior introduced by configuration changes
  before production deployment.
%  before rolling the changes to production.

As discussed in \S\ref{sec:validation}, existing configuration validation
  is segregated from the code using configurations, and can hardly
  cover all the constraints or deal with bugs exposed by configuration changes.
Our prior work, PCheck~\cite{xu:16}, explores the feasibility of using
  the code from the original software to check configuration values.
Despite the promising results, we have come to the conclusion that PCheck's method
  is fundamentally limited.

First, PCheck is significantly incomplete due to its difficulty
  in dealing with external dependencies
  and avoiding side effects.
PCheck only detects around 70\%
  of the well-scoped misconfigurations~\cite{xu:16} (all of them can be
  exposed by configuration testing, \S\ref{sec:transform}).
Second, PCheck identifies misconfigurations solely based
  on generic error signals (exceptions, error code, program termination).
It cannot deal with semantic errors or undesired behavior,
  and thus cannot combat legal misconfigurations.

Configuration testing addresses the above limitations.
It can exercise code with side effects;
  external dependencies can be mocked or auto-generated (cf.~\S\ref{sec:creation}).
Legal misconfigurations can be captured by asserting expected behavior,
  as how assertions are used in software tests.
% The configuration-as-code practice has already setup
%  culture and
%  infrastructure support for configuration testing.

% automate~\cite{detreville:05,Leuschner:2017}

% Treating configuration as code naturally eliminates
% the boundary between software engineers
% and sysadmins. It fundamentally changes the landscape of configuration management by
% breaking a few longstanding assumptions:
% (1) sysadmins do not understand the system implementation;
% (2) sysadmins do not have access to source code;
% and (3) sysadmins do not read or write code.
% In essence, misconfigurations no longer lie in
%  the gray zone between developers and sysadmins~\cite{xu:13,yin:11};
% instead,
%  engineers take full responsibility in reviewing, validating
%  configuration changes, and actively enhancing validation
%  as postmortems to configuration related incidents.

\section{Concluding Remarks}

% \section{Conclusion}

To conclude,  this paper presents the proposal of
  configuration testing as a key reliability
  engineering discipline
  for configuration management in large-scale production systems.
The essence of treating configuration as code is to apply
  rigorous software engineering principles and techniques
  for configuration management,
  which should go beyond current practices.
We hope that this paper will open the direction of configuration testing
  and inspire innovations and endeavor to make testing a
  regular practice for system configuration.

\section*{Acknowledgement}

We thank Darko Marinov, Peter O'Hearn,
  Alex Gyori, and David Chou
  for the invaluable discussions on the idea
  and the early drafts
  of configuration testing.

\footnotesize
\bibliographystyle{acm}
\bibliography{ref}

\end{document}